# ANALYSIS OF TRAVEL ACTIVITY DETERMINANTS USING ROBUST STATISTICS

**Václav Plevka**[*]
L-Mob Leuven Mobility Research Centre, CIB
KU Leuven
Celestijnenlaan 300A - PO Box 2422, 3001 Leuven, Belgium
0032 16372846
vaclav.plevka@kuleuven.be

**Pieter Segaert**
Department of Mathematics
KU Leuven
Celestijnenlaan 200B – PO Box 2400, 3001 Leuven, Belgium
0032 16372337
pieter.segaert@kuleuven.be

**Chris M. J. Tampère**
L-Mob Leuven Mobility Research Centre, CIB
KU Leuven
Celestijnenlaan 300A - PO Box 2422, 3001 Leuven, Belgium
chris.tampere@kuleuven.be

**Mia Hubert**
Department of Mathematics
KU Leuven
Celestijnenlaan 200B – PO Box 2400, 3001 Leuven, Belgium
0032 16322023
mia.hubert@kuleuven.be

## ACKNOWLEDGEMENTS

The Authors would like to thank the Federal Public Planning Service Science Policy (Belspo) and Dr. Eric Cornelis of FUNDP Namur for providing us with the dataset, and Vlaamse Overheid (IWT-SBO project "Urban Logistics and Mobility" - 140433) and Vlaams Instituut voor Mobiliteit (project P081 "Transmob", IWT ref 140203) for providing funding. Mia Hubert and Pieter Segaert acknowledge grant C16/15/068 of the Internal Fund KU Leuven.

---

[*] Corresponding author

Vaclav Plevka, Pieter Segaert, Chris M. J. Tampère, Mia Hubert

**ABSTRACT**

This study investigates travel behavior determinants based on a multiday travel survey conducted in the region of Ghent, Belgium. Due to the limited data reliability of the data sample and the influence of outliers exerted on classical principal component analysis (CPCA), robust principal component analysis (ROBPCA) is employed in order to reveal the explanatory variables responsible for most of the variability. Interpretation of the results is eased by utilizing robust principal component analysis (ROSPCA).

The application of ROSPCA reveals six distinct principal components where each is determined by a few variables. Among others, our results suggest a key role of variable categories such as journey purpose-related impedance and journey inherent constraints. Surprisingly, the variables associated with journey timing turn out to be less important.

Finally, our findings reveal the critical role of outliers in travel behavior analysis. This suggests that a systematic understanding of how outliers contribute to observed mobility behavior patterns, as derived from travel surveys, is needed. In this regard, the proposed methods serve for processing raw data typically used in activity-based modelling.

# 1        INTRODUCTION

This paper reports on innovative techniques to reveal key travel behavior determinants of a sample population. The method utilizes travel survey data conducted on a regional scale. The methodological contribution stems from employing specific statistical techniques that allow processing multidimensional data reliably. As with most travel survey data sets, inherent to our data are outlying observations with "questionable" values.

## 1.1      Motivation

A logical prerequisite for operationalizing any activity-based model is understanding the target population. This knowledge is obtained from travel surveys that contain comprehensive information about conducted activities such as when and where the activities were executed or which travel mode was chosen. Here, every person is characterized by a unique multidimensional mobility pattern that in theory consists of all their trips. It is convenient if this pattern is summarized in a limited set of selective and distinctive determinants, such that personal mobility behavior can be sufficiently distinguished from that of others. Loosely speaking, such a set of determinants could be called a personal 'mobility finger print'. However, to the best of our efforts, we were able to find relatively few contributions in literature defining such finger print, or providing methodologies to reduce big (longitudinal) data sets to summarized mobility pattern summary variables.

The use of big data, which allows professionals to acquire seemingly complete meta-data regarding spatial and temporal individual characteristics, is rapidly emerging in the transport domain. The challenge associated with their processing is even more pertinent: more methods that allow practitioners to explore and utilize the full potential of a data source effectively are required in the field of travel behavior studies. To process these data sets various multivariate statistical methods might be conveniently used. A very effective statistical technique in this context is principal component analysis or PCA. While reviewing related literature (in the next section: *Literature study*), it has turned out that systematic understanding of how data quality affects the resulting analysis is still limited. Note that data quality addresses to the presence of outliers in multivariate data space.

This paper demonstrates the impact of outliers by comparing classical and robust PCA. Both techniques are applied to the same household travel survey data which include variables such as travel activity timing (e.g. number of peak time journeys), length (e.g. mean work-journey distance), frequency (e.g. number of education trips) or inherent constraint (e.g. number of journeys with baggage). However robust PCA inherits the interpretability issue of classical PCA. The resulting loading vectors are typically comprised of many of the original variables. This issue may be addressed by classical sparse principal component analysis but this procedure is highly sensitive to possible outliers in the data. Therefore we employ sparse robust PCA in order to improve the interpretation of the exploratory analysis. As a result, most principal components are dominated by just one variable. In the next section we discuss current literature demonstrating the role of PCA in travel behavior research. As a final section of the introduction we demonstrate the effects outliers may have on even the simplest statistical procedures. Finally we show the need for specific procedures call robust statistical techniques which can handle multivariate outliers.

## 1.2      Literature study: a need of dimension reduction in travel behavior research

Dimension reduction techniques, a family of multivariate statistics, have been employed in transport research in different contexts; this section only highlights the applications to travel behavior analysis.

The need for dimensional reduction stems from the inherent complexity of travel behavior. For example, our original data set (see detailed description in *Data resource*) consists of 139 travel activity variables for every record. Therefore, it is viable to employ some dimension reduction technique that narrows the scope of the analysis to the most determining characteristics. Several authors already proposed to use PCA to reduce the complexity of the data in travel behavior research. For example, (Joly 2004) employed PCA to construct 5 principal components out of 14 transport-urbanistic variables. Consequently, the components were used to explain the dependencies between the urban structures and travel time budget that quantifies the space and time accessibility. (Steg 2005) projected 33 car-attractiveness indicators on 3 principal components that capture motives of car use. (Van and Fujii 2007) used 16 belief variables on transport modes characteristics collected from 208 surveyees in Ho Chi Minh City to determine 3 principal components which were identified as Symbolic Affective, Instrumental and Social Orderliness. The components were used to classify attitudes towards different travel modes in the city. (Hunecke et al. 2007) projected 21 psychological variables on 8 principal components that capture attitude towards ecologic use of transport means. (Sohn and Yun 2009) examined differences between car-dependent (using a car regardless the alternative options) and normal-choice commuters and analyzed a role of both groups in a mode choice analysis. PCA resulted in 6 factors using 19 psychometric variables on car-usage motives which were collected by an online survey. The factors were later used as independent variables in a mode choice. (Pitombo and Gomes 2014) studied work travel behavior of workers in São Paulo Metropolitan Area. The 3 resulting principal components, classified as socio-economic class, urban environment and family structure, were built by PCA using 22 variables collected by a travel survey. One of several results was the proof of dependency between the workers socio-economic background, or family structure and commuting characteristics. Finally, (Gim 2015) projected 26 land use and "life situation" variables on 3 principal components that depict automobile-travel utility. The aforementioned examples well demonstrates that the principal components derived from the original data might be conveniently used for explanatory or modelling purposes. Another popular application of PCA in travel behavior analysis is in market segmentation which ensures that a transport policy tailors to specific traveler's needs. A thorough discussion of different market segmentation techniques is provided in (Wedel and Kamakura 2000). In essence, meaningful user groups (market segments) are constructed according to key consumer characteristics. In contrast to a priori, say expert-based segmentation, PCA reveals the important travel behavior characteristics in a systematic, data-driven approach. For example, (Anable 2005) analyzed the interdependence between personal attitude and travel behavior, specifically on transport mode orientation. In total 105 attitudinal statements were subjected to PCA that resulted in 17 significant factors. This allowed construction of 6 distinct market segments representing the respondent's car-orientation. (Wittwer 2014) used 15 travel activity variables such as travel activity frequency, timing or duration to construct 8 principal components that allowed the segmentation of young adults in Germany into 6 distinct groups. In (Kandt et al. 2015), 63 attitudinal statements were transformed with regard to the respondent's relation to sustainable and information technologies into 13 principal components that allowed to build 6 mobility profiles for each traveler in the investigated regions.

The previous examples clearly illustrate PCA is a valuable well-established statistical technique for travel behavior research. However the reliability of travel surveys is inevitably burdened by the presence of outliers. Their impact was already documented in (Talvitie and Kirshner 1978; Witlox 2007; Van Acker and Witlox 2010; Stigell and Schantz 2011; Singleton 2013). Systematic treatment of contaminated multivariate data however has received very little attention. Analyzing travel activities specifically in the spatial context, (Buliung and Remmel 2008; Schönfelder and Axhausen 2003) tested a number of statistical tools explicitly suited to spatial analysis that might facilitate single attribute outliers identification in space. (Chen et al. 2008) proposed a robust Mahalanobis distance-based algorithm capable of identifying the spatial multiple-attribute outliers. However, analysis of travel activity determinants is considerably more complex as it encompasses both discrete and continuous variables on time, space and travel characteristics. (Jin et al. 2008) used robust statistics as a preprocessing step to reduce human effort in finding potential loop detector faults in traffic flow pattern studies. (Singleton 2013) demonstrated the identification of outliers by means of robust statistics, but focused on combining univariate robust techniques applied to each variable. The dependence structure between the variables of the multivariate observations is therefore lost. In the next section we will illustrate that a univariate outlier detection approach for multivariate data is insufficient and a multivariate outlier detection approach is paramount. Moreover the procedure of (Singleton 2013) still requires extensive manual intervention by the practitioner.

Recent developments in travel behavior analysis and modelling - for reference see (Van Acker et al. 2016) - emphasized "multidimensionality" of the travel behavior concept that encapsulates mode choice, travel frequency, route and destination choice, journey scheduling and timing, trip chaining or transport resources ownership decisions, etc. These decisions are also influenced by level of comfort, attitudinal traits, daytime, household (family) arrangements or neighborhood properties. In addition, it is likely that with a growing amount of substantially more complex data sets collected by emerging techniques and technologies (Axhausen 2008; Hasan et al. 2013; MacFarlane 2014; Wu et al. 2014; Soora 2014; Picornell et al. 2015; Huang and Wong 2016) and with demand for deeper insights from these novel data resources, the role of outliers becomes critical. With it the need for robust techniques becomes even greater.

## 1.3    The need for robust multivariate statistics

The following discussion outlines some issues associated with applying dimension reduction techniques that might arise if data sets are of limited reliability. PCA is based on correlations among variables; therefore its successful application is vulnerable to the presence of multivariate outliers in the data. Outliers may be described as atypical observations deviating from the pattern suggested by the majority of the data. A variety of methods, for example histograms or boxplots, might be employed for screening each variable separately and consequently for detecting univariate outliers. However, difficulties with those methods arise if the investigated data are spread over many dimensions. To illustrate, we consider a subset of the bike sharing data discussed in (Fanaee-T and Gama 2013). The data consists of hourly and daily data of the number of bike trips by registered and non-registered users in the Washington D.C. area for the years 2011 and 2012. The full data is available at the UCI Machine Learning repository (Lichman 2013). Figure 1 shows a scatterplot of the daily number of trips taken by registered users (horizontal axis) and the total number of trips (vertical axis). The grey points correspond with measurements taken on working days, while the blue points correspond with non-working days. The latter observations are clearly deviating from the working days, although both their x-coordinate and y-coordinate are in line with the values observed on the non-working days. Consequently, univariate outlier detection methods, such as the boxplots plotted along the axes, are unable to flag any unusual observation. On this plot we have also drawn the 99.5% classical tolerance ellipse with dashed border. It contains all points whose squared Mahalanobis distance is smaller than the 99.5%-quantile of the chi-squared distribution with 2 degrees of freedom. If all data were following a bivariate normal distribution, we would expect only 0.5% of them to fall outside of this ellipse. As the Mahalanobis distance is based on the classical mean and covariance matrix, the resulting tolerance ellipse is inflated by the outlying non-working days and does not detect all of the non-working days as outliers. The ellipse with full border is tighter and does separate the working days from the non-working days. This ellipse is based on a robust estimator of center and scatter, the Minimum Covariance Determinant (MCD), introduced by (Rousseeuw 1984). Robust statistical methods have been developed to overcome a possible impact of outliers. We notice that the MCD also discovers that some of the working days have unusual values. Both November 25, 2011 and November 11, 2012 correspond to Black Friday, which implies very specific travel behavior. The same happened on March 23, 2012 as there was a National Cherry Blossom Festival in de Washington DC area. Not taking into account these outliers in subsequent analyses might distort the conclusions. Whereas in two dimensions one might still inspect the data for these points using a bivariate plot, there is no direct exploratory analogue for higher dimensional data. The techniques presented in the methodology section are robust in the sense that they provide similar results when there are outliers in the data as when there are no outliers in the data. Moreover no user intervention is needed. Particularly, robust dimension reduction techniques are applied on a multidimensional travel survey data set. A comparison of robust and non-robust methods clearly demonstrates the critical influence of outliers which would mislead any interpretation of travel behavior.

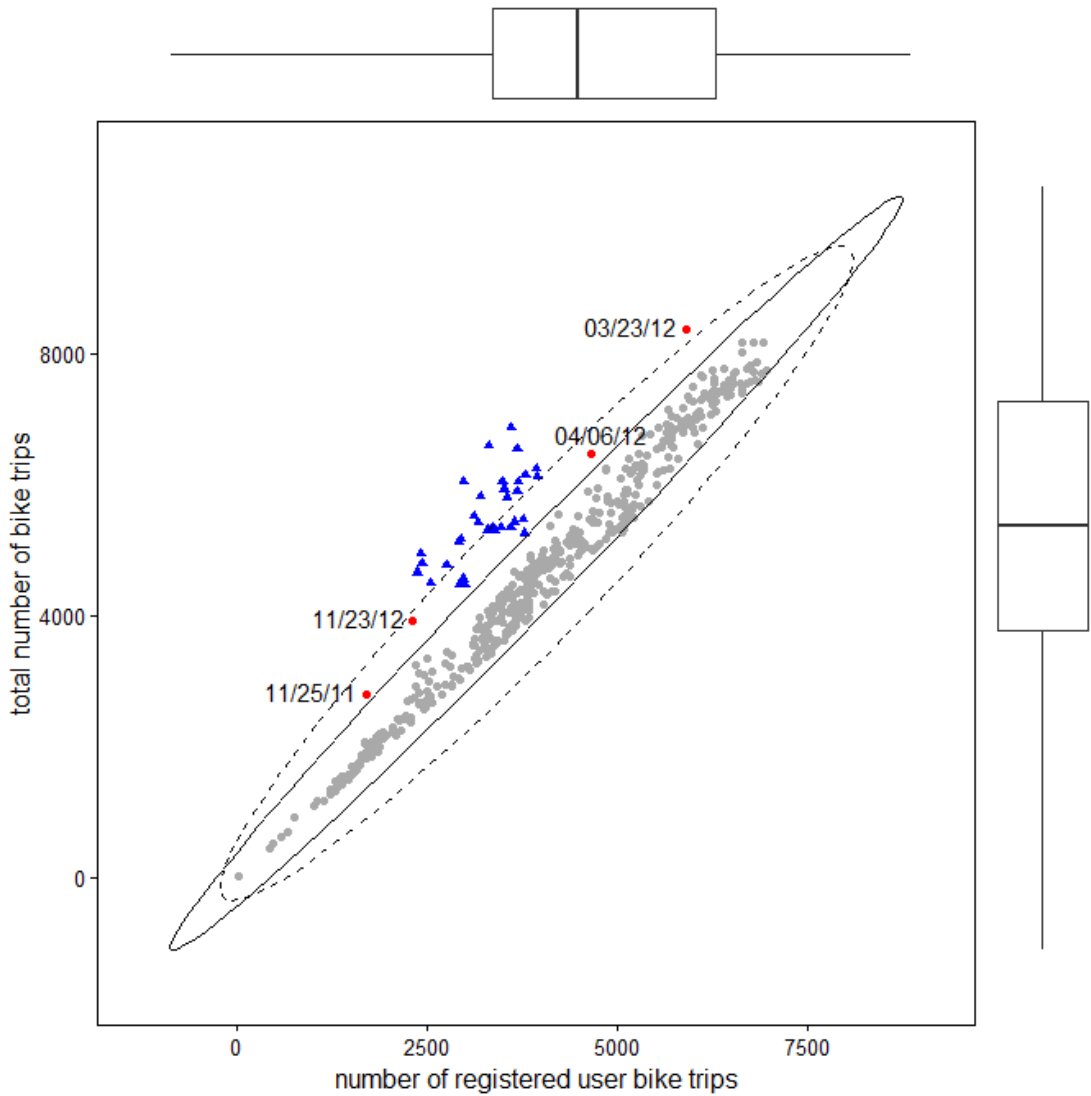

**< Figure 1 Multivariate outliers in two-dimensional space, with univariate boxplots and classical and robust tolerance ellipse superimposed >**

## 2      METHODOLOGY

Classical principal component analysis (CPCA) aims to find linear combinations of the original variables capturing most of the covariance structure of the original data. These combinations are chosen such that they are orthogonal and sequentially maximize the variance of the data projected on them. Let $X \in \mathbb{R}^{n \times p}$ be a data matrix with $n$ observations of dimension $p$. CPCA can be described as a search for a center $\hat{\mu}$ and unit-norm orthogonal loading vectors, the columns of $P$, such that the resulting scores defined as $(X - 1_n \hat{\mu}')P$ have maximal variance. Here, $1_n$ is the vector of length $n$ containing 1's. More formally the $j$-th column of $P$ can be defined as the unit norm vector maximizing $S\left(p_j'(x_1 - \hat{\mu}), \dots, p_j'(x_n - \hat{\mu})\right)$ under the constraint that $p_j$ is orthogonal to the previous $j - 1$ vectors Here $S$ denotes the standard deviation and $\hat{\mu}$ is the mean of $X$. The loading vectors one then obtains coincide with the eigenvectors of the classical covariance matrix of $X$. Typically one does not retain all the directions, but rather chooses the first $k$ components explaining a sufficient amount of the total variance of the original data. A very popular criterion is based on the *scree plot* which plots the sorted decreasing eigenvalues versus their index. The number of directions corresponding to the point at which an elbow occurs is then selected. Another simple but popular criterion is to retain the first $k$ principal components such that at least *a predefined percentage* of the total variance is explained by these first $k$ directions.

However CPCA often suffers from interpretability difficulties as the elements of the loading vectors are typically neither very small nor very large in absolute value. To overcome this issue classical sparse principal components analysis (CSPCA) has been proposed by (Jolliffe et al. 2003; Zou et al. 2006). Those methods yield loading vectors $P$ with many zero values, which makes it easier to interpret the principal components. In (Jolliffe et al. 2003) it is proposed to incorporate a non-sparsity penalty parameter and thus to maximize $S\left(p_j'(x_1 - \hat{\mu}), \dots, p_j'(x_n - \hat{\mu})\right) - \lambda\|p_j\|$ instead. A higher value of $\lambda$ corresponds to greater sparsity and for $\lambda = 0$ CPCA is recovered. A Bayesian Information Criterion (BIC) may be used to select the optimal value of the sparsity parameter $\lambda$.

A second main disadvantage of CPCA, also shared by CSPCA, is its dependence on classical estimators of variance which are highly sensitive to the possible presence of outliers. These outliers may have various effects including changing the retained principal components both in number and direction. Robust alternatives of CPCA have been proposed by various authors including (Croux and Ruiz-Gazen 2005; Hubert et al. 2005; Hubert et al. 2002). We will focus on the ROBPCA approach by (Hubert et al. 2005) as a robust alternative for CPCA and the recent robust sparse PCA method called ROSPCA by (Hubert et al. 2015). The latter approach combines sparseness and robust PCA ideas. We will now briefly sketch both the ROBPCA and ROSPCA approach.

ROBPCA combines two important concepts in robust statistics: projection pursuit and robust covariance estimation. Projection pursuit techniques are based on the insight that outlying observations should deviate from the bulk of the sample in at least one direction. For example, the non-working days in Figure 1 cannot be detected when projecting the data onto the coordinate axes, but they stand out on the $-45°$ direction. In the first stage of the ROBPCA algorithm the data are projected onto many directions (hence the name projection pursuit). On each direction a robust measure of location and scale is used to measure the outlyingness of the data points along that specific direction. A multivariate measure of outlyingness of each point is obtained by considering its maximum outlyingness over all considered directions. An initial robust subspace of dimension $k$ is obtained by finding the (classical) principal components based on the $h$ (with $0.5n \leq h \leq n$) observations with smallest outlyingness. Next, CPCA is applied to all observations that are close enough to this robust subspace of dimension $k$. Depending on the number $h$ and the number of outliers and their position, this reweighted subspace might be based on (many) more observations than $h$. Note that $h$ constitutes a lower bound on the regular number of observations. Without any knowledge about the possible amount of contamination, it is taken as $h = 0.5n$ which corresponds to assuming that at least half of the data points are not outlying. In the last stage of ROBPCA all points are projected onto this reweighted subspace and the MCD estimator is used to obtain a robust measure of location and a robust scatter matrix. The eigenvectors of this robust scatter matrix then determine the final robust loadings and corresponding eigenvalues. For more details see (Hubert et al. 2005). The ROBPCA procedure is available in the free statistical software package R (R Core Team 2013) and in LIBRA (Verboven and Hubert 2005), a free Matlab library for robust statistics.

Note that ROBPCA is a highly robust method which can withstand any proportion of outliers up to $(n - h)/n$ , no matter where they are located with respect to the outlier-free group. Some outliers may be scattered around, while others could be clustered in one or several groups. Many other methods, such as robust probabilistic PCA, do not satisfy this property and make a distributional assumption about the regular points and the outliers.

ROSPCA is comparable to ROBPCA but it leads to sparse loading vectors. Again, in the first step projection pursuit techniques are used to find a set of $h$ observations with the smallest degree of multivariate outlyingness. CSPCA is then applied to this subset. As in the ROBPCA approach a reweighting step is performed in which all observations are considered that are close to the sparse subspace obtained in the previous step. Finally a robust measure of scale is employed within the subspace to obtain the final eigenvalues whilst keeping the sparsity structure of the loadings. Similar to the CSPCA case a BIC criterion may be used to select the sparsity parameter. An implementation of the ROSPCA algorithm for R is available at http://wis.kuleuven.be/stat/robust and will be freely available on CRAN shortly.

Both robust procedures also allow us to create a diagnostic plot proposed in (Hubert et al. 2005). For each of the observations two distances are calculated: the score distance and the orthogonal distance. The score distance may be seen as a robust Mahalanobis distance of the scores. It thus measures how far an observation lies from the rest of the data within the PCA subspace. By contrast, the orthogonal distance of an observation equals the Euclidean distance of the observation to its projection into the PCA subspace. It therefore measures how far a point lies from the PCA subspace. Based on these two distances, one may categorize observations into regular observations; orthogonal outliers (with large orthogonal distance); good leverage points (with small orthogonal distance and large score distance); and bad leverage points with both high orthogonal and score distance. Appropriate cutoff values to separate the different type of observations are derived in (Hubert et al. 2005) and (Hubert et al. 2009). Here, we follow the method of (Hubert et al. 2009) and define the cutoffs as the upper whiskers of the adjusted boxplot computed on the respective distances. The adjusted boxplot was introduced in (Hubert and Vandervieren 2008) and has whiskers which depend on the skewness of the data. At asymmetric data this yields a better separation between regular observations and outliers than the standard boxplot does.

## 3        PCA ANALYSIS: GHENT USE CASE

The data input comprises data collected as a part of the project Behavior and Mobility within the Week (BMW) that was undertaken as a common initiative of KU Leuven and the University of Namur (Viti et al. 2010). The BMW project conducted a comprehensive travel behavior analysis of Ghent region, Belgium. In total, 717 individuals recorded all travel activities that were carried out over a 7-day course, observed in a period between September and December 2008. The collected data can be split into two databases: first, the socio-demographic database that contains information about the personal and household background, for example age, address, household size etc., and second, the travel activity (TA) database that contains revealed preferences data about 19471 journeys. Figure 2 (a), Figure 2 (b) and Figure 2 (c) present the distributions of Total number of journeys, Mean journey distance and Total number of peak-time journeys, respectively for the different occupation categories as they were defined in the survey. The figures outline some relevant insights into travel patterns in Ghent as a university city: while students and schoolchildren represent the second most frequently travelling group with 16.5 % from the total amount of observed journeys and with 18.5 % from the total amount of peak journeys, their typical travel journey distance is below the average in comparison with other occupation groups in the sample. On average they do the second shortest trips. Noteworthy, the student and schoolchildren group accounts for 18.1% of the total sample size.

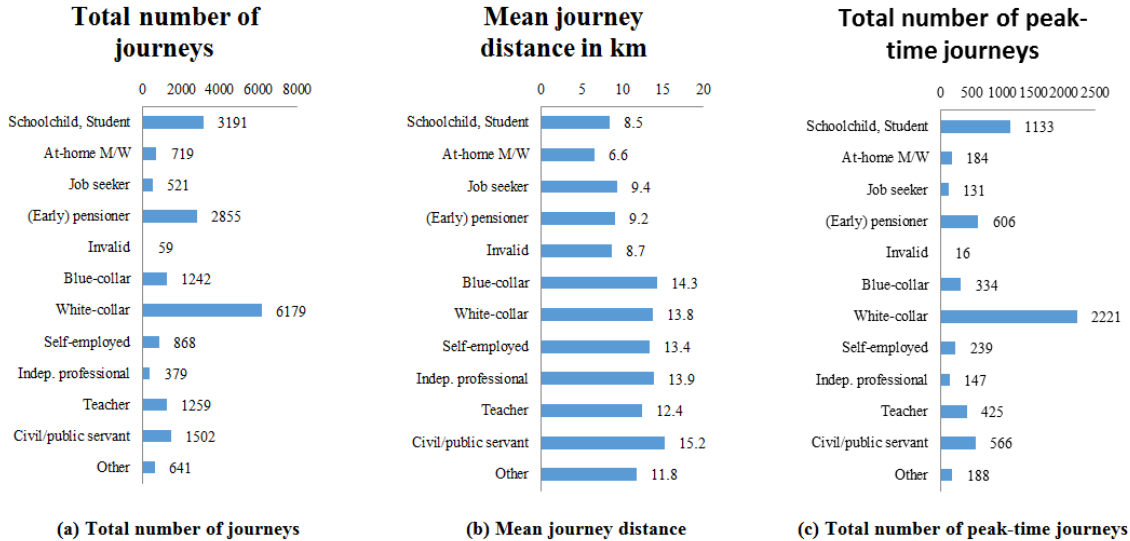

(a) Total number of journeys    (b) Mean journey distance    (c) Total number of peak-time journeys

< Figure 2  Fundamental travel characteristics by the different occupation categories >

For the travel activity determinants analysis presented in this study, the socio-demographic database was withdrawn and focus was devoted exclusively to the TA data. In contrast to other studies of mobility behavior utilizing PCA, the variables containing transport mode related information were disregarded here. This constraint was imposed due to our initial motivation: the analysis was intended to predict modal choice and vehicle ownership, so no modal choice information should be present in the exogenous variables. Furthermore, there were some technical requirements which imposed additional data restrictions. For example, categorical non-ordinal variables can be only very hardly processed by PCA.

Vaclav Plevka, Pieter Segaert, Chris M. J. Tampère, Mia Hubert

**Table 1 Essential statistics for TA variables which were used as an input for the analysis**

| Variable | Unit | Mean | StDev | Median | MAD |
|---|---|---|---|---|---|
| Number of days with no journey | - | 0.444 | 0.850 | 0.000 | 0.000 |
| Number of Peak-time [7-9, 16-19] journeys | - | 8.633 | 5.156 | 8.000 | 3.000 |
| Number of Night time journeys | - | 1.478 | 2.561 | 1.000 | 1.000 |
| Number of Escort (bring/pick up someone) journeys | - | 2.130 | 3.475 | 0.000 | 0.000 |
| Number of To home journeys | - | 10.266 | 3.757 | 10.000 | 2.000 |
| Number of Work journeys | - | 2.759 | 3.043 | 2.000 | 2.000 |
| Number of Education journeys | - | 0.990 | 2.074 | 0.000 | 0.000 |
| Number of Shopping (go eat, daily shopping, regular shopping) journeys | - | 4.117 | 3.153 | 4.000 | 2.000 |
| Number of Leisure (personal business, family/friend visit+sport/culture/touring) journeys | - | 6.815 | 4.517 | 6.000 | 3.000 |
| Mean escort-journey distance | km | 4.375 | 10.823 | 0.000 | 0.000 |
| Mean to home-journey distance | km | 11.756 | 11.308 | 8.052 | 4.348 |
| Mean work-journey distance | km | 11.530 | 23.034 | 2.060 | 2.060 |
| Mean education-journey distance | km | 2.419 | 8.613 | 0.000 | 0.000 |
| Mean shopping-journey distance | km | 5.694 | 10.469 | 3.000 | 2.000 |
| Mean leisure-journey distance | km | 13.195 | 19.199 | 8.056 | 4.764 |
| Number of short duration trips (<=5min) | - | 17.452 | 9.822 | 6.000 | 4.000 |
| Number of short distance trips (<=1km) | - | 9.709 | 8.041 | 4.000 | 4.000 |
| Mean journey distance of the most frequent activity | km | 11.655 | 11.218 | 5.067 | 3.710 |
| Number of journeys with children | - | 3.844 | 6.733 | 0.000 | 0.000 |
| Number of journeys with purchased goods | - | 3.749 | 3.554 | 3.000 | 2.000 |
| Number of journeys with baggage | - | 5.643 | 7.442 | 2.000 | 2.000 |

Table 1 presents an overview of the TA variables; for each TA variable, the measurement units and four basic data statistics are displayed. Note that the median and the median absolute deviation (MAD) represent robust measurements of location and scale respectively. The lower value of the median compared to the mean indicates that all variables are right skewed. Zero values of the median and the MAD simultaneously imply that half of the measurements are equal to zero. At some variables we notice a large difference between the classical and the robust scales. This hints at the possible presence of outliers advocating the use of ROBPCA. Additionally, the variables were assigned to five distinct categories according to some meaningful context they provide. Some reported TA purposes were aggregated on a level desirable for this study (e.g. the variables ta8 and ta9). The variables ta16 and ta17 were derived from the BMW dataset by measuring duration and distance, respectively of the most frequent origin – destination relation, withdrawing all "To-home" type of activities. Thus, the variables ta16 and ta17 might serve as an indicator of household allocation with respect to a place of primary interest.

## 4    PCA ANALYSIS: RESULTS

As the data consist of several variables with different units and scales, the data are first centered and scaled using the mean and the standard deviation. The scree plot for the classical PCA analysis is shown in Figure 3 (a). The percentage of variance explained by the first $k$ components is shown for the first seven components. The first five principal components are retained as those explain at least 80% of the variance. The corresponding loadings are shown in Figure 4 (a).

The five components are mostly comprised of multiple variables and nearly every variable contributes to at least one principal component. To inspect the data for possible leverage points and outliers, the CPCA diagnostic plot is shown in Figure 5.

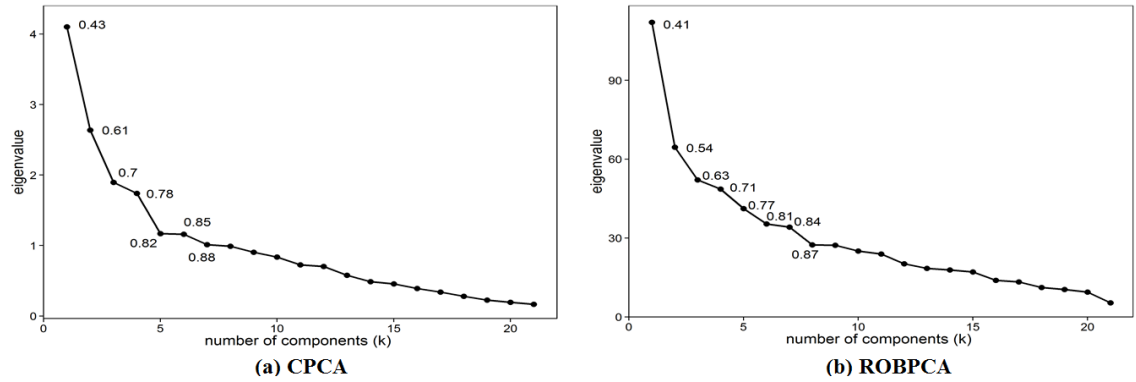

**(a) CPCA**  **(b) ROBPCA**

**< Figure 3 Scree plots showing the amount of explained variance by CPCA and ROBPCA >**

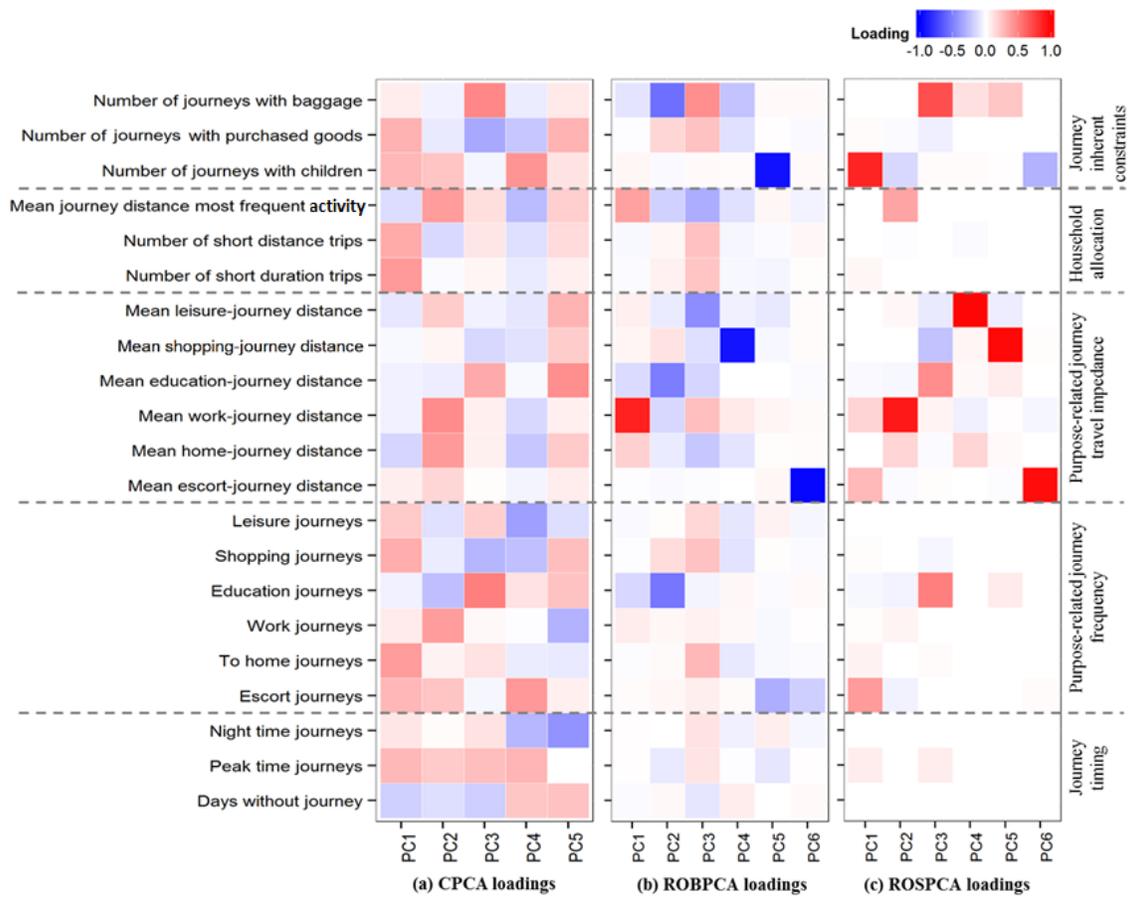

**< Figure 4 Overview of PCA results: Variable loadings per component for CPCA, ROBPCA and ROSPCA >**

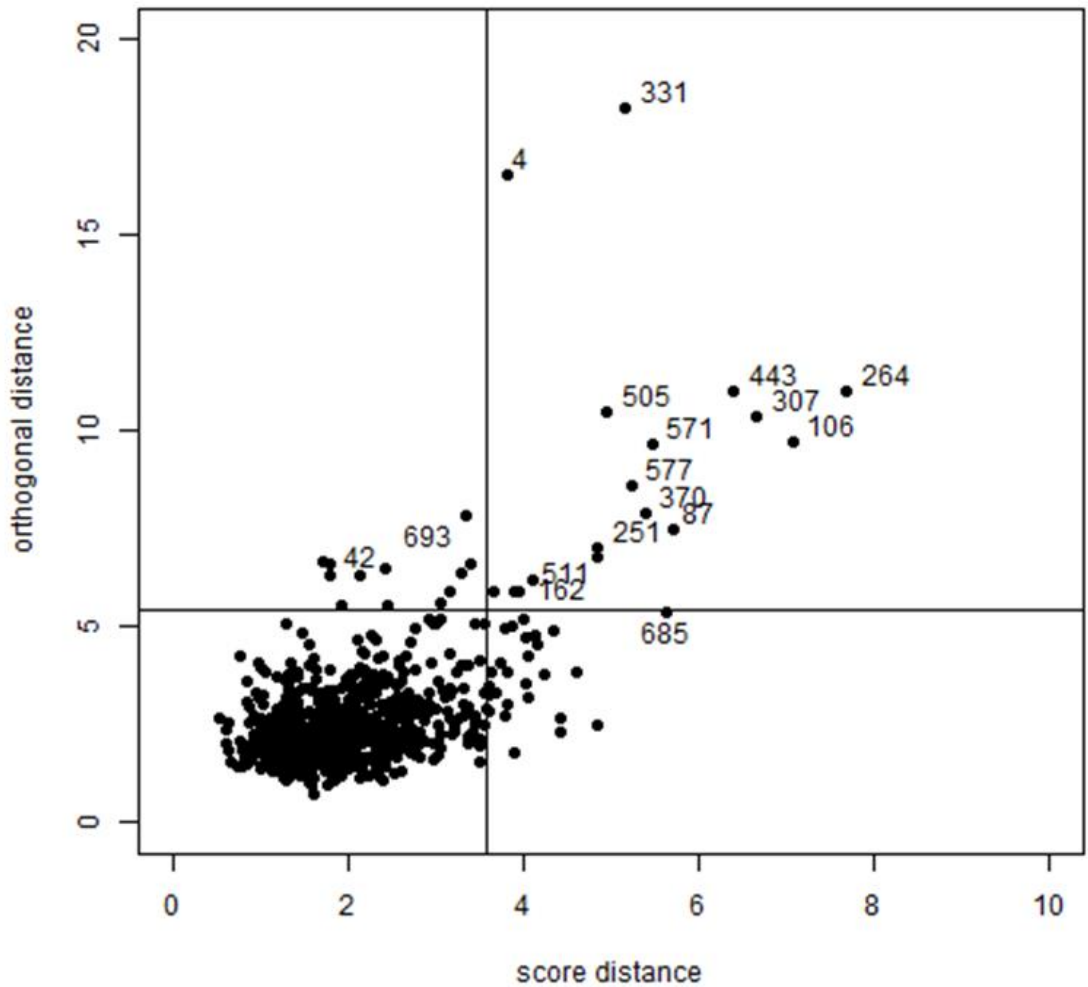

**< Figure 5 CPCA diagnostic plot >**

Most striking are observations 4 and 331 with the highest orthogonal distance and observations 264, 106, 307 and 443 with the highest scores distances. Inspection proved these observations were indeed deviating from the bulk of the sample. Observation 4 has a mean escort distance of 185 km, whereas the mean shopping distance of observation 331 is 206 km. These values are well above typical values found in the data set. The lower loadings of the corresponding variables for the 5 PCA vectors explain the high orthogonal distance. Similar findings can be made for the group of leverage points which were found to have exceptionally large values for several variables, including the mean distance of the most frequent journey and the mean education, work or home distance. The sparse classical PCA analysis is very similar, hence the results are not shown here. The BIC criterion of CSPCA selected a sparsity parameter of only 0.075, so almost no penalization was performed.

However, the ROBPCA analysis suggests a different view on the data. Again the data are first standardized. The centering was carried out using the median. Variables for which the MAD was nonzero were scaled using the MAD. Figure 3 (b) shows that the scree plot now forces the inclusion of an additional sixth principal component to explain at least 80% of the total variance.

The loadings of the robust principal components, see Figure 4 (b), are far more concentrated on some of the variables, with several of the variables having relatively small contributions to the six principal components. To aid the interpretation of the principal components we also apply the ROSPCA procedure described above. Using the BIC criterion a sparsity parameter of 0.23 was selected. The obtained loadings from the ROSPCA procedure are shown in Figure 4 (c). Comparing these loadings to the loadings obtained from ROBPCA shows that the variables which contribute mostly to each principal component are still the same. However, variables with small contributions to the ROBPCA components are now penalized, resulting in many zero loading values. The resulting diagnostic plot is shown in Figure 6. It is very similar to the diagnostic plot of ROBPCA, which is not shown here.

First we notice that the scales of both the score distances and the orthogonal distances have significantly increased. This indicates that the corresponding distance measure for the CPCA analysis was severely affected by these atypical observations. The most important bad leverage points for CPCA no longer have a highly deviating score distance in the ROSPCA analysis. Several observations, such as 42, 162 and 693, clearly stand out in the robust analysis; comparatively, in the classical analysis these points can rather be considered boundary cases and may not be flagged as outliers. Inspection revealed that these observations have high values for variables such as mean escort, work and shopping distance. The high importance of these variables in the PCA subspace explains their high score distance.

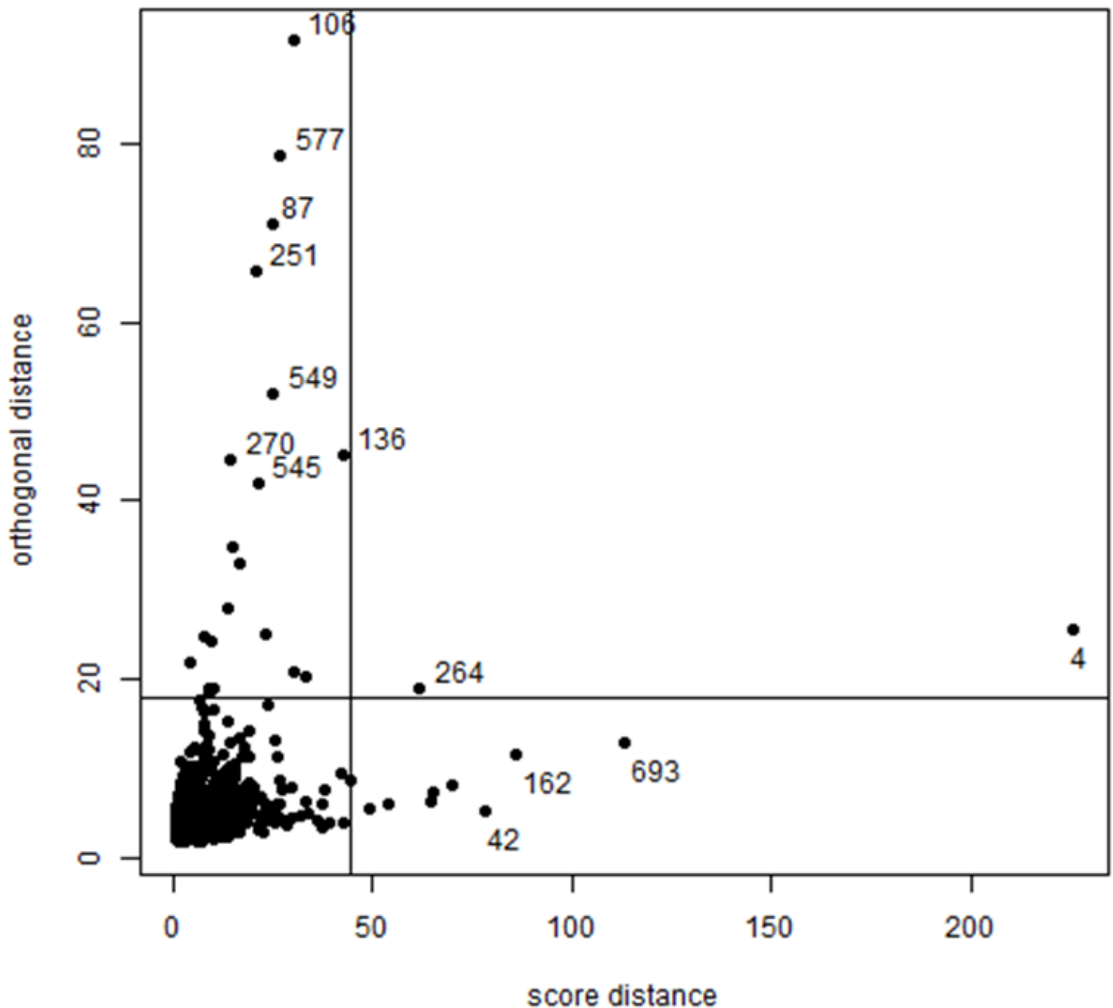

**< Figure 6 ROSPCA diagnostic plot >**

The previous remarks clearly indicate that the classical analysis is heavily influenced by the outlying observations. They not only pull the PCA subspace in a different direction, resulting in principal components made up of different variables, but also result in a different number of principal components and make some clear outliers appear to be regular points. In the robust statistics literature, this phenomenon is known as 'masking'. The benefit of robust procedures and the interpretability benefit of robust sparse principal component analysis are clearly shown.

## 5        INTERPRETATION OF ROSPCA

The following subsections provide behavior interpretation of the ROSPCA results evaluating the PC loadings presented in Figure 4 (c). Convenience of the chosen method is evident from visual inspection: the resulting components are "dominated" by one variable that is accompanied by few other, noticeably less-loaded variables. This loading pattern considerably eases the interpretation of the PCA results. By screening the patterns in Figure 4 (c), it can be stated that the category `Purpose-related travelling impedance' and `Journey inherent constraints' play a decisive role in the travel behavior of the sample under study.

Figure 7 provides a visualization of the results using the Gephi SW package. The figure depicts component loading through color-coding (red shades for positive loadings and blue shades for negative loadings) the edges connecting the nodes, which here represent PCs and TAs. Similarly, each variable category from Table 1 is designated by a different color. Additionally, the size of the nodes representing PCs indicates the amount of explained variance. Thus, the first principal component PC1 is portrayed as the largest circle, and so on. For simplicity, only the loadings higher or equal to an absolute value of 0.1 are visualized.

Among others, Figure 7 confirms a decisive role of the variable categories Purpose-related journey travel impedance and Journey inherent constraints. On the other hand, since the connections with very low loadings were withdrawn, the journey timing variables dropped-out of consideration. Moreover, limited contribution from the remaining variable categories (Purpose-related journey frequency and Household allocation) is apparent as all the related nodes have just one connecting edge with the highest loading equal to 0.5.

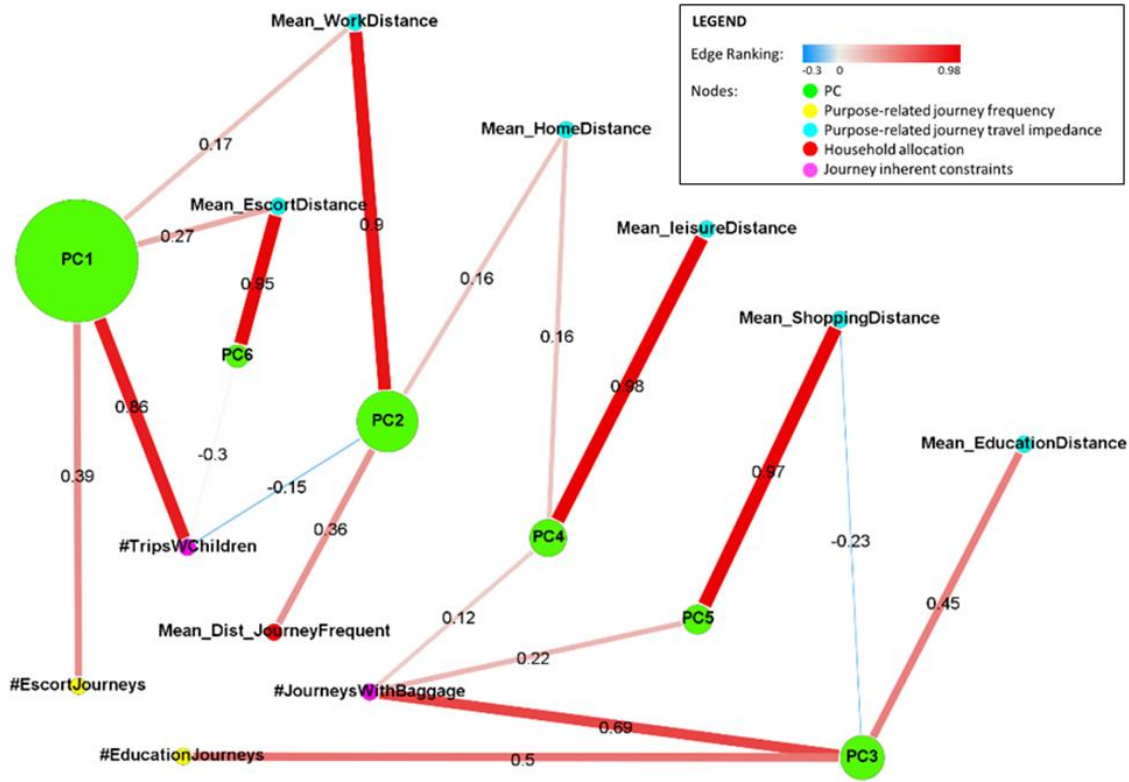

**< Figure 7 Network interpretation of the ROSPCA results depicting the relations between principal components and loaded variables >**

## 5.1 PC1: Traveling with kids

The first principal component explains most of the variance contained in the TA data. In that sense, the first principal component is the most important one. The highest loaded variables on this PC are (sorted in decreasing order of PC loadings absolute values):

- ta19: Number of journeys with children,
- ta4: Number of Escort journeys
- ta10: Mean Escort-journey distance and
- ta12: Mean Work journey distance.

Not surprisingly, the variable ta19 that is mostly loaded here is accompanied by the escort-type variables ta4 and ta10. For example, a simple tour whose primary purpose is to fetch kids from school would be composed of (at least): (1) an actual journey with kids and (2) an escort trip heading to school. Although this study does not attempt to segment population, the first PC could be interpreted as an indicator of travel behavior of a "(working) parents" user group.

## 5.2 PC2: Working

The second most important principal component is loaded mostly by the following variables:

- ta12: Mean Work journey distance;
- ta18: Mean journey distance of the most frequent activity; and
- ta19: Number of journeys with children.

This component is determined by the impedances of "obligatory" travel activity ta12. It might be hypothesized that the obligatory journeys are recursive and additionally play a decisive role in deciding household location (Brueckner 2011). In this light, it can be expected that ta12 is accompanied by the highly loaded ta18. Interestingly, ta19 is loaded with a negative sign while it is highly positively loaded onto PC1 (this functional link is clearly observable in Figure 7). PC2 would virtually split the population sample according travel behavior into two disjointed groups: users doing work-related journeys, and users traveling with children.

### 5.3    PC3: Education traveling with considered constraints

The third principal component is loaded mostly by the following variables:
- ta21: Number of journeys with baggage;
- ta7: Number of Education journeys; and
- ta13: Mean Education-journey distance.

Next to the work related TAs, it is expected that the Education TAs play an important role in describing the sample's travel behavior because Ghent is a university city with most students having the habit of traveling to their parents' home for weekends; this explains why education related variables and journeys with baggage are combined in this PC. Moreover, as ta21 is the most loaded explanatory variable, the travel-inherent constraints such as travelling with baggage, are decisive for this kind of travel behavior.

### 5.4    PC4: Leisure

The fourth principal component is loaded mostly by the following variables:
- ta15: Mean leisure-journey distance;
- ta11: Mean Home journey distance; and
- ta21: Number of journeys with baggage

PC4 is dominated (one variable is loaded much more than others) by the leisure-related travel impedance. The contribution by ta21 suggests that the journey-inherent travel constraints are also perceived during leisure travel.

### 5.5    PC5: Shopping

The fifth principal component is loaded mostly by the following variables:
- ta14: Mean Shopping-journey distance; and
- ta21: Number of journeys with baggage.

Similarly to PC4, this component is dominated by a non-obligatory journey-type impedance variable: Shopping-related travel impedance. The high loading of ta21 might be explained by travelling with purchased goods, here indicated as "baggage" rather than "goods" (ta20, here with zero loading).

### 5.6    PC6: Escort journeys

The last principal component that explains the least amount of the first six components is loaded mostly by the following variables:
- ta10: Mean Escort-journey distance; and
- ta19: Number of journeys with children.

Unlike PC1, which accounts for trips with children and Escort journeys, here the component attempts to explain Escort journeys without kids. In other words, PC6 captures Escort journeys impedance (positive sign) while the frequency of journeys with children is loaded with negative sign.

## 6    CONCLUSION

The presented study provides additional evidence of the chosen methodologic feasibility in the context of travel behavior research. We were able to demonstrate clearly that the results from classical PCA analysis were heavily influenced by outliers. In fact, the outliers were so dominant that the two variable categories were no longer detectable in the resulting PCs. In contrast with CPCA, ROSPCA delivered PC loadings that allow for easily drawn conclusions due to functionally distinct PC loadings, as evident in Figure 7. Moreover, employing ROSPCA allows for remarkably good behavioral interpretation of the PCs.

Utilizing ROSPCA on longitudinal survey results of the city of Ghent in Belgium, the travel behavior determinants were successfully revealed. The results showed that the PCs are predominantly loaded by two variable categories: (1) journey purpose-related travel impedances and (2) journey inherent constraints. The importance of the first variable category is well known and the travel impedance standardly serves as a utility component in transport modeling. The determining role of second variable category represented by *ta19: Number of journeys with children* and *ta21: Number of journeys with baggage* is intuitive: travelling with (for example) children directly influences travel modality (it is likely a car mode), timing (bringing kids to school at morning) and other travel characteristics. On the other hand, it is somewhat surprising that no considerable contribution has been noted from the category Journey timing variables. In other words, what distinguishes the travel behavior of examined population is highly dominated by travel constraints (kids, luggage) in addition to impedance variables of regular trips and trip purpose. A possible explanation for this might be that the activity scheduling is less important for time-flexible user groups, such as self-employed, independent professionals, job seekers, stay-at-home men/women and particularly in the case of Ghent, students. Figure 2 hinted at the typical travel patterns of students that might be characterized be relatively frequent but short-distance journeys. Although the analysis of travel modes was purposely neglected, it can also be assumed that most of the trips by students and schoolchildren are done by non-motorized modes independent of peak hour phenomenon.

The authors remain unsure about an exact role of ta21 at PC5 due to its ambiguous definition. The original survey did not further specify the difference between terms *baggage* and *purchased goods*. Both terms provide interchangeable interpretation. This ambiguity could explain why ta21, instead of the expected *ta20: Number of journeys with purchased goods,* is highly loaded here. These results allow us to "finger print" the population sample based on analysis of their entire mobility pattern. In future, the presented statistical methods may considerably facilitate the data processing for travel activity analysis same as the actual analysis, particularly in the cases when novel (multidimensional) data resources enabled by the emerging technologies are used.